\documentstyle[preprint,aps]{revtex}

\tightenlines

\begin{document}

\title{Measuring Neutron Separation Energies Far from Stability}

\author{W.A. Friedman$^a$ and M.B. Tsang$^b$}

\author{\small $^a$ Department of Physics, University of Wisconsin,
Madison, WI 53706, USA}

\author{\small $^b$ National Superconducting Cyclotron Laboratory and Department
of Physics and Astronomy, Michigan State University, East Lansing,
MI 48824, USA}

\date{\today}

\maketitle

\begin{abstract}
A method is proposed for the experimental measurement of neutron
separation energies for nuclei far from stability. The procedure
is based on determining cross sections for the production of
nuclei, by projectile fragmentation, for which only protons are
removed but for which the number of neutrons is left unchanged. A
simple Abrasion-Ablation analysis leads to a cross section
prediction which is sensitive to the neutron separation energy
after a single parameter is adjusted in comparison with data.
Examples which illustrate the method are presented.
\end{abstract}


Few experimentally measured values of neutron separation energies
have been determined for nuclei near the neutron drip-line
\cite{aw1, aw2, aw3, ganiln}.
However, it is precisely in this
region that such information is particularly necessary for
investigating nucleosynthesis, and for testing models of nuclear
structure at the greatest distance from the valley of stability
\cite{val,val1, val2}. As the values of $N$ and $Z$ move away from
the valley of stability on the neutron-rich side, the neutron
separation-energy systematically is reduced \cite{val,val1, val2}
due primarily to the asymmetry in the proton-neutron composition.
This asymmetry is reflected by the value of the parameter $\delta
$ which is defined as $(N-Z)/(N+Z)$. The separation energies for
nuclei with large asymmetry are often extrapolated from
information available close to the valley of stability
\cite{acong}. This approach may be uncertain in describing how
their values goes to zero
with increasing $%
\delta $. However, this region tests the role of the symmetry energy most
stringently in theoretical models, and hence provides information about the
symmetry energy in the more general context of the nuclear equation-of-state
\cite{pav}. Thus, the systematic behavior of the separation energy
with increasing values of $%
\delta $ offers direct evidence for such effects.

One of the most successful methods for the production of neutron-rich rare
nuclides has been the fast projectile fragmentation process \cite{ben,fri}.
This method has been used to extend the list of particle stable nuclei to
the extremes \cite{ca2, camore}.
In most of the cases where
the values of the neutron separation-energy have been measured, the values
are in excess of $5$ MeV. In the case of nuclides very close to the
drip-line
the pairing
energy may cause 2-neutron separation energies to be less than the 1-neutron
separation energies. For those cases the lowest separation energy, referred
to as $S$ in this article, be it for one or two neutrons, is the one of
interest.
In this paper we suggest that,
under certain circumstances, the same measurement which provides the
verification for the existence of a rare nucleus may also be used to
estimate $S$.

One avenue to this information lies in the recent suggestion that
\textquotedblleft cold\textquotedblright\ fast fragmentation \cite{ben,fri}
seems an efficient method of producing these extremely rare nuclei. The
simple scenario for this process follows the concepts of the
Abrasion-Ablation (A-A) models \cite{gs,ben,fri}. From that point of view, a
direct reaction (abrasion) removes a number of nucleons, leaving the residue
excited, and free to lose more nucleons by evaporation (ablation).

To minimize the uncertainties associated with the abrasion and ablation
processes, we focus on the production of fragments where only protons and no
neutrons are removed from the projectile, i.e., the abrasion process removes
the protons, and leaves the residue with too little excitation energy
to permit further loss of particles (neutrons).
Such a
production mechanism is referred as a ``p-removal chain'' in this article.
If the residue is neutron-rich, it decays by neutron emission. Thus the
upper limit on the excitation energy is the neutron separation-energy, $S$.
It is this feature that provides the production cross section with the
sensitivity to the separation energy. By limiting
our attention to the nuclei produced in p-removal chains, we avoid the
ambiguities related to specific evaporation models.

Using the frame work of the A-A model \cite{fri}, the cross section to
produce a nucleus with $(Z-x)$ protons and $N$ neutrons from the
fragmentation of a projectile with $Z$ proton and $N$ neutrons can be
written as $\sigma _{x}=Abr_{x}\cdot Abl_{x}$. Here, the factor $Abr_{x}$
gives the cross section for removing $x$ protons (and no neutrons) by
abrasion, and $Abl_{x}$ is the dimensionless probability that the residue
will not further decay following the removal of those $x$ protons.

The factor $Abr_{x}$ can be estimated by the geometric overlap \cite{gos} of
projectile with the target. This can provide the cross section for the
removal of x particles \cite{gs,ben}. This cross section must then be
multiplied by the probability that all the abraded particles are protons.
By assuming that the positions of neutrons and protons in the projectile are
uncorrelated, the probability that all $x$ of the abraded particles are
protons can be written as $(Z!/(Z-x)!)/((N+Z)!/(N+Z-x)!)$. Both of these
assumptions (geometric overlap and uncorrelated positions) are simplistic
but their validity can be calibrated by comparison with a measured set of
cross sections for projectile fragmentation where the separation energies of
the resulting nuclei are known.

The crucial factor is $Abl_{x}$ which depends on both the distribution of
excitation energy following the removal of $x$ protons, $F_{x}(E^{\ast })$,
and the separation energy, $S_{x}$, for the nucleus produced by this
removal. Specifically, $Abl_{x}$ is the integral of the excitation function
from zero to the separation-energy.

Clearly the form of the excitation function is a critical component of
this procedure. The literature of A-A models suggests that this distribution
function may be quite uncertain \cite{gs,ben}. Recent approaches \cite{ben, gs}
suggest that the distribution function, $F_{x}(E^{\ast })$, is a convolution
of $x$ distribution functions, $f_{1}(e^{\ast })$, where $f_{1}(e^{\ast })$
is the function for the removal of a single nucleon:
\begin{equation}
F_{x}(E^{\ast })=\int \prod_{i=1}^{x}(de_{i}^{\ast }\;f_{1}(e_{i}^{\ast
}))\;\delta (\sum_{i=1}^{x}e_{i}^{\ast }-E^{\ast })
\end{equation}%
The functional form of $f_{1}(e^{\ast })$ is, however, not well determined,
and there is great uncertainty as to the mean value of the excitation
energy, $<e^{\ast }>$, it provides. Some of this uncertainty can be removed
by fitting calculated cross sections to sets of measured cross sections. The
fitting would be accomplished by the adjustment of
$<e^{\ast }>$.

One form of the suggested single particle excitation distribution, $%
f_{1}(e^{\ast })$, widely used \cite{gs} is the \textquotedblleft
triangle\textquotedblright\ distribution. This has the form $f_{1}(e^{\ast
})=2/E_{m}(1-e^{\ast }/E_{m})$ for $e^{\ast }<E_{m}$ and the average
excitation energy, $<e^{\ast }>$ is $E_{m}/3$. A wide range values of $E_{m}$
have been suggested in different models \cite{gs, ben}. Convolution of this
(triangle) single-particle distribution leads to a value for $Abl_{x}$ which
is approximately $(2S_{x}/(3<e^{\ast }>))^{x}/x!$, for $S_{x}<<3<e^{\ast }>$%
. Exact values for $Abl_{x}$ can be obtained with
\begin{equation}
Abl_{x}=C_{tri}(x)\cdot (2S_{x}/(3<e^{\ast }>))^{x}/x!,
\end{equation}%
where
\begin{equation}
C_{tri}(x)=\sum_{s=0}^{x-s}(-S_{x}/(3<e^{\ast }>))^{s}\cdot
(x!^{2}/(s!(x+s)!(x-s)!))
\end{equation}%
The value of $C_{tri}(x)$ goes to 1.0 for small values of $%
(S_{x}/(3<e^{\ast }>))$, and $Abl_{x}$ is seen to be a function of the
parameter $(2S_{x}/(3<e^{\ast }>))$.

We have also considered a different form for the single-particle excitation
function, i.e., the exponential function $f_1(e^{\ast })=1/<e^{\ast }>%
\mathrm{exp}(-e^{\ast }/<e^*>)$ with an average excitation energy $<e^{\ast
}>$. A convolution of this function provides an $x$-particle distribution
function of the form
\begin{equation}
F_{x}(E^{\ast })=(E^{\ast }/<e^*>)^{x-1}/(x-1)!\mathrm{exp}(-E^{\ast }/<e^*>)
\end{equation}%
When this function is integrated from zero to the separation energy $S_{x}$,
one obtains a value for $Abl_{x}$ which is approximately $%
(S_{x}/<e^*>)^{x}/x! $, and an exact expression can be calculated as a
function of $(S_{x}/<e^{\ast }>)$:
\begin{equation}
Abl_{x}=C_{exp}(x)\cdot (S_{x}/<e^{\ast }>)^{x}/x!,
\end{equation}%
with
\begin{equation}
C_{exp}(x)=\sum_{s=0}(-S_{x}/<e^{\ast }>)^{s}(x/((x+s)s!))
\end{equation}%
The value of $C_{exp}(x)$ goes to 1.0 for small values of $%
(S_{x}/<e^{\ast }>)$.

For small values of $S_{x}$, the functional form of
the $Abl_{x}$ (as a function of different parameters) is the same for the
two single-particle distribution functions. For the \textquotedblleft
triangle\textquotedblright\ distribution the parameter is $%
(2S_{x}/(3<e^{\ast }>))$, while for the exponential distribution the
parameter is $(S_{x}/<e^{\ast }>)$. Hence one might expect that a choice of $%
<e^{\ast }>$ in the \textquotedblleft triangle\textquotedblright\
distribution which is 2/3 the choice of$<e^{\ast }>$ in the exponential
distribution would predict similar cross sections. When the full $x$%
-particle excitation distributions are used, however, this scaling is only
approximate.

For the nuclei in a given p-removal chain, estimates of unknown values of $%
S_{x}$ can be extracted from measured values of cross sections, $\sigma_{x}$,
after a value
of $<e^{\ast }>$ has been adjusted to fit the data for nuclei with known
separation energies.

In illustrative examples below we have found that we can, generally, well
represent the data by first calculating $Abr_{x}$ using simple assumptions,
and then adjusting the parameter $<e^{\ast }>$ to give $Abl_{x}$. The
assumptions for $Abr_{x}$ include both the estimation of the removal cross
section from the geometric-overlap of target and projectile, and also the
calculation of probability for obtaining pure proton removal by assuming
uncorrelated positions for the nucleons. Any
systematic correction required in $Abr_{x}$ may possibly be accommodated by
adjusting $<e^{\ast }>$ in $Abl_{x}$. The
ambiguity in the choice of the single particle distribution function, $f_{1}$%
, prevents a unique determination of the mean excitation energy through the
fitting. However, we do find interesting systematic changes in the required
values of $<e^{\ast }>$ which appear to depend on the mass and isospin
values of the fragmenting projectile.

We have considered, for illustration, the p-removal chains given in reference%
\cite{ben} for $^{208}Pb$. $^{{197}}Au$ and $^{136}Xe$. We have also
examined data in the literature for $^{86}Kr$ \cite{kr}, and $^{48}Ca$ \cite%
{ca2}, and, in addition, preliminary results for $^{58}Ni$ which is under
current investigation \cite{tsang}. We first tested the form for the cross
section suggested in the expressions of Eqs. 2 and 5, by fitting to data for
the fragmentation production of nuclei where the separation energies are
known. Both the \textquotedblleft triangle\textquotedblright , and the
exponential forms for the excitation functions were used, and fits to the
data were achieved by adjustment of the respective values of $<e^{\ast }>$.

In Fig. 1, we plot the fitted cross sections for the fragmentation
of $^{86}Kr$. The separation energies \cite{ganiln} are known for
all first 5 members the p-removal chain (the data only covers
2-5). There are 3 degrees of freedom for this fit providing a
$\chi ^{2}$ per degree of freedom of .99 for the exponential
distribution. The respective fitting values of $<e^{\ast }>$ are
11.7 MeV for the \textquotedblleft triangle\textquotedblright\
distribution, and 16.6 MeV for the exponential distribution.

In Table I we
list the values of the $\chi ^{2}$/degree of freedom for fits to other data
sets including only nuclei where the separation energies are
known \cite{ganiln}. Excellent fits can be achieved.
We have also listed values of the fitting parameters $<e^*>$
with estimates of deviations (both plus and minus) for 70\% confidence.
There are
variations in the values of these fitting parameters from one reaction to the
another. In each case an approximate ratio of approximately 2:3 is found for
the values related to the
\textquotedblleft triangle\textquotedblright\ and the exponential $%
f_{1}(e^{\ast })$ distributions with practically no
difference in the resulting fits from the two distributions.
Except for the $Xe$ fragmentation, the data and calculated values,
based on the excitation energies
which provide the
best fit \cite{fit}, are plotted in Figures 1 to 4.
Preliminary
results show similar behavior for the p-removal chain of $^{58}Ni$ for which
the chain has been measured through 8 p-removals \cite{tsang}.

While the current data is quite limited we also examined the
predictive power of the method for nuclei where the separation
energy is unknown, even lacking estimation by extrapolation. For
this we looked at $^{204}Pt$ which was measured in the chain from
$^{{208}}Pb$ projectile \cite{pb}. In Fig.3 we show the best fit
to the data for the first three nuclei in the chain, where
separation energies are known or estimated \cite{ganiln}. In the
inserted graph we show the sensitivity to the assumed binding
energies of the $^{204}Pt$ using the values of $<e^{\ast }>$ which
best fit the first three members of the chain given in Table I.
The apparent estimate for the separation energy is about 5 MeV
with large uncertainty due to the experimental uncertainty in the
measured cross section for $^{204}Pt$. This value is well in line
with systematic decrease of the separation energy with increased
$(N-Z)$.

We have also attempted to estimate the separation energy of
$^{41}Al$ which has recently been observed \cite{ca2} for the
first time. Even though the fragmentation of neutron-rich
$^{48}Ca$ \cite{ca2,camore,ca1} has been studied intensely in the
past few years, there is no systematic measurement of cross
sections for the p-removal chain. The separation energies are not
known for the nuclei with more than four protons removed. Some of
the extrapolated values have uncertainties of much more than 1 MeV
\cite{ganiln, camore}. Thus in principle the $^{48}Ca$ would be a
good place to fully examine our method. We have examined the
existing data to make a rough estimate. We found the fragmentation
cross sections have been measured from previous studies \cite{ca1}
for two nuclei, $^{45}Cl$ and $^{44}Si$, which have 3 and 4
protons removed from $^{48}Ca$. This experiment used a target of
$^{9}Be$. Using the separation energies 6.241 MeV and 5.21 MeV
\cite{ganiln, camore} respectively for $^{45}Cl$ and $^{44}Si$ the
fit parameters listed in Table I are obtained. We next examined
the results from the experiment \cite{ca2} which first observed
the $^{41}Al$ nucleus corresponding to the removal of 7 protons
from the projectile. Unfortunately, these data were obtained with
a $^{181}Ta$ target. To connect this point with the other two
points (obtained with a Be target), we used abrasion calculations
which suggest that the difference in targets provides a cross
section from Be which is 0.545 times the value obtained with a Ta
target. The reduction arises primarily from the difference in the
respective size of the impact parameters for the two targets. We
plot in Fig. 4
a point for $x=7$ ($^{41}Al$) at a cross section of $4.4$ pb%
, which is the value of $8$ pb \cite{ca2, saku}, reported for the
Ta target, scaled down by the estimated ratio of cross sections.
(We have also scaled down the error bar.) The insert in Fig. 4
shows the dependence of the calculated cross section for $x=7$ as
a function of the separation energy. The
experimental uncertainty is high since only three events were observed \cite%
{ca2, saku}. Even so, extraction of a value of $3.5\pm 0.5$ MeV
would be consistent with the information in the inserted graph. We can not
claim that this value is, indeed, the separation energy of $^{41}Al$ due to
the fact that two different targets were used, and there is a scarcity of
information in the p-removal chain
(For example, there are no measured cross-sections and no accurate separation
energies for for x=5 and 6 ($^{43}P$ and $^{42}Si$) isotopes).
However, this exercise shows the potential for
extracting the separation energies for $^{44}S$, $^{43}P$, $^{42}Si$, as
well as $^{41}Al$, the nuclei with 4, 5, 6, and 7 protons removed from
$^{48}Ca$. This might be accomplished with careful measurements of the complete
p-removal chain from x=1 to 7 with one target and one beam energy.
Specifically, additional data in the x=1-3 region where the $S_{x}$ values
are known by observation will provide greater constraints on the values of $%
<e^{\ast }>$.

Finally, we have examined the situation for the fragmentation of
$^{197}Au$ where a chain of 5 proton removal is reported in ref.
\cite{ben} for a target of $^{9}Be$ (solid points in Fig. 5. For
the fragmentation of the $^{197}Au$ using an $^{27}Al$ target
nuclei, there are p-removal cross sections up to x=3
\cite{ben,khs}. For the case of $^{197}Au$ projectiles with
$^{27}Al$ and $^{9}Be$ targets, abrasion estimates suggest a 5\%
reduction in going from the larger to the smaller target
\cite{friben}. The three open points in Fig. 5 are the $^{27}Al$
data scaled down by 5\%. They are consistently higher than the
corresponding $^{9}Be$ data (solid points). If we apply the
fitting procedures to this set of data, using the values of
$<e^{\ast }>$ listed in Table I, we obtain a separation energy
greater than $7$ MeV for both $^{193}Re$,  and  $^{192}W$ nuclei.
These values are clearly inconsistent with systematic trends and
expectations, both of which would have led to values below 7.0
Mev. For comparisions, the solid and dashed lines are calculations
using the best fit $<e^{\ast }>$ listed in Table I for $^{9}Be$
(36.3 MeV) and for $^{27}Al$ (32.2 MeV) targets, with the
assumption of an exponential energy distribution. The upper and
lower curve in each pair of lines use the separation energy of 7.0
and 6.5 MeV respectively as the separation energy for both
$^{193}Re$ and $^{192}W$ nuclei. The calculated cross-sections are
lower than the experimental values. In brief, the reported cross
sections for the $^{197}Au+^{9}Be$ reaction do not lead to
reasonable separation energies for the last two members of the
proton chain. The reasons for this failure are not clear at this
time.

In summary, the illustrated calculations show that excellent agreement with
fragment cross-sections, where the fragment neutron separation energies are
known, can be obtained when the simple estimates are used with the
Abrasion-Ablation model, and a single parameter, $<e^{\ast }>$, is adjusted.
The quality of agreement is equally good for both the \textquotedblleft
triangle\textquotedblright\ and the exponential
single particle excitation distribution functions.
The two distributions require values of
the mean energy, which are approximately in the ratio of 2:3.
The smaller the parameter the slower the fall of cross section
with the number of protons removed. The quality of the fit inspires
confidence in the use of the A-A model for calculating the p-removal chains.
Once the parameter is determined for each chain the only remaining input is
the set of separation energies. From some of the data in the literature we
were able to suggest the power of the p-removal method for observing unknown
separation energies in $^{204}Pt$ and $^{41}Al$. A puzzling disagreement was
found for the unknown separation energies of $^{193}Re$ and $^{192}W$ in the
chain reported for the fragmentation of $^{197}Au$ when the target was
$^{9}Be$. Clearly, more data and more understanding of the uncertainties in
the cross-section measurements are needed to confirm the utility of the
method. Since the procedure may be generally applied to all p-removal
chains, it opens an avenue for measuring separation energies for neutron
rich nuclei near the drip-line as illustrated by the fragmentation of
extremely neutron-rich projectiles such as $^{48}Ca$. While the method can
not compete with dedicated mass measurements where masses can be measured
to uncertainties better than $10^{-7}$ \cite{val,val1,val2}, the simplicity of
cross section measurements with fragment separators may allow the wide use
of this method to measure the separation energies for extremely neutron rich
nuclei to a couple hundred keV as this energy decreases toward zero at the
drip-line.

This work was supported in part by grants from the US National Science
Foundation, PHY-0070161 and PHY-01-10253.

\newpage



\newpage


\begin{table}[t]
\caption{ Values of $<e^*>$
are obtained by fitting the measured cross sections
of nuclides with know separation
energies. The columns labelled + and - indicated the deviations of the
the best fit values with 70\% confidence.
The 2$^{nd}$ and 6$^{th}$ column reflect the goodness of the
the fit for the ``triangle'' (tri) and exponential (exp) distributions.}

\vspace{0.2cm}
\begin{center}
\footnotesize
\begin{tabular}{|c|c|c|c|c|c|c|c|c|}
\hline
Reaction&$\chi^2/dof$&$<e^*>$&-&+&$\chi^2/dof$&
$<e^*>$&-&+\\
&tri.&tri.&&&exp.&exp.&&\\
\hline
$^{208}Pb+Cu$\cite{pb}&0.38& 18.4&1.1&1.5&0.42&26.6&1.8&{2.1}\\
$^{197}Au+^{27}Al$\cite{khs}&0.87&22.4&1.6&3.6&0.88&32.2&3.8&{5.2}\\
$^{197}Au+^{9}Be$\cite{ben}&1.87&25.0&1.4&1.8&1.58&36.3&2.2&{2.6}\\
$^{136}Xe+^{9}Be$\cite{khs}&0.36&23.8&2.8&2.6&0.36&34.2&3.8&{5.6}\\
$^{86}Kr+^{9}Be$\cite{kr}&1.45&11.7&0.3&0.25&0.99&16.6&0.4&{0.45}\\
$^{48}Ca+^{9}Be$\cite{ca1}&1.24&7.70&0.35&0.4&1.81&10.80&0.45&{0.60}\\
\hline
\end{tabular}
\end{center}
\end{table}

\newpage

{\bf FIGURE CAPTIONS:}

Figure 1. Measured cross sections (solid circles) for the
production of p-removal nuclides with $N=50$, $^{86-x}Z$ from the
fragmentation of $^{86}Kr$ with a target of $^{9}Be$ \cite{kr}.
Lines are predictions described in text using two different
excitation distributions with the adjustment of a single
parameter, given in Table I.

Figure 2. Measured cross sections for the production of p-removal
nuclides with $N=126$,$^{208-x}Z$ from fragmentation of of
$^{208}Pb$ with a target of $^{63}Cu$ \cite{pb}. Lines are
predictions described in text using two different excitation
distributions with the adjustment of a single parameter, given in
Table I. The insert shows the predicted cross-sections as a
function of the separation energy for $^{204}Pt$ nuclei. The
horizontal solid and dashed lines are measured cross-sections.

Figure 3. Measured Cross sections for the production of p-removal
nuclides with $N=28$,$^{48-x}Z$ from fragmenation of $^{48}Ca$
with $^{9}Be$ target \cite{ca1} (solid points). Lines are
predictions for x=1,2,3,4 described in text using two different
excitation distributions with the adjustment of a single
parameter, given in Table I. The open point for x=7 ($^{41}Al$) is
obtained from separate experiment with $^{181 }Ta$ target
\cite{ca2} and adjusted as describe in the text. The insert shows
the predicted cross-sections as a function of the separation
energy for $^{41}Al$ nuclei. The horizontal solid and dashed lines
are measured cross-sections.

Figure 4. Measured cross sections for the production of p-removal
nuclides with $N=118$,$^{197-x}Z$ from a projectile of $^{197}Au$
with $^{9}Be$ target \cite{ben} (solid points). The open circle
are data \cite{khs} for $^{197}Au+^{27}Al$ scaled down by 0.95.
Lines are predictions described in the text.


\begin{thebibliography}{99}
\bibitem{fit} A variation of 10\% in $<e^*>$ will change the calculated
cross section for the removal of x protons by approximately $x\cdot10$\%
\end{thebibliography}
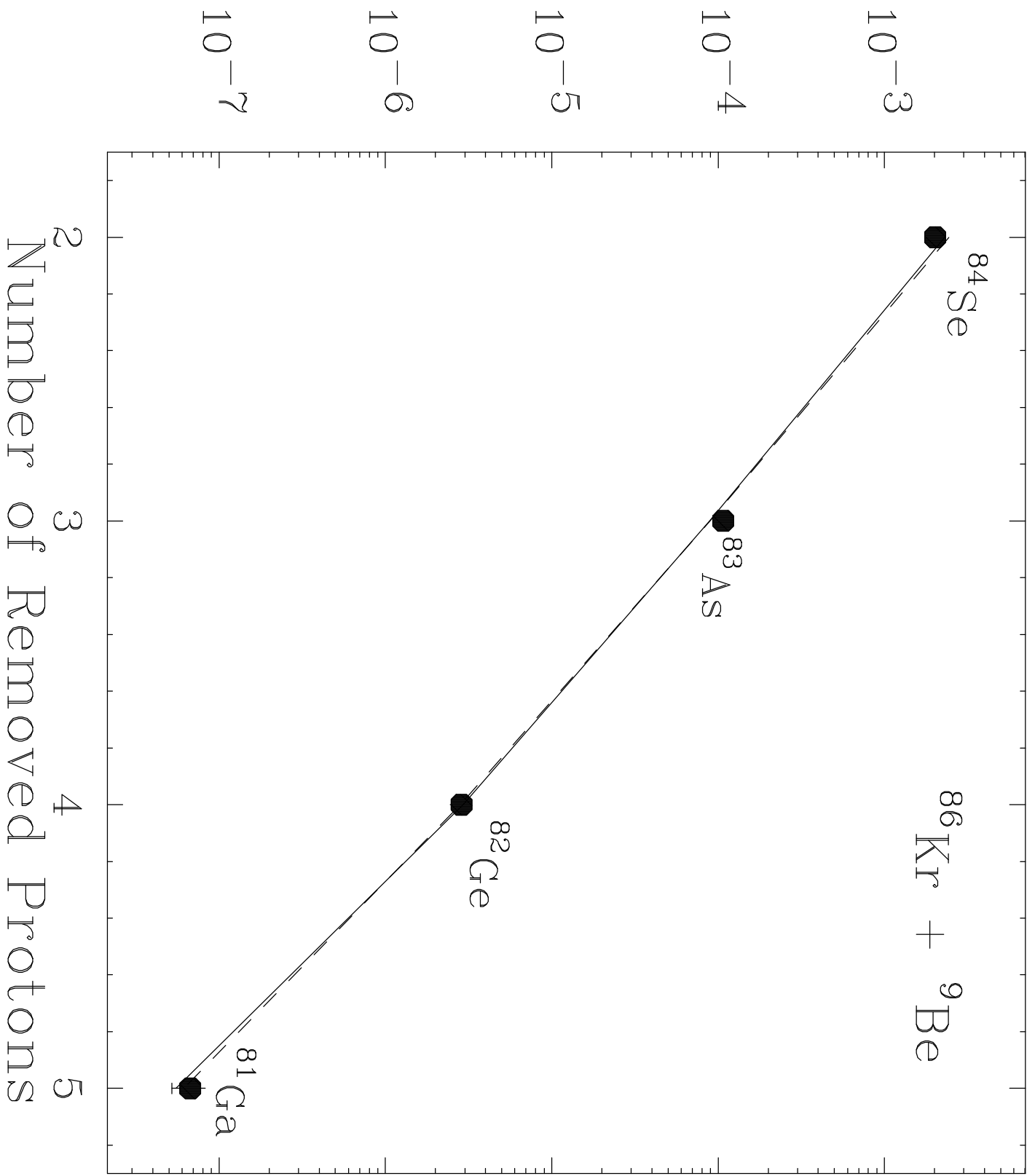
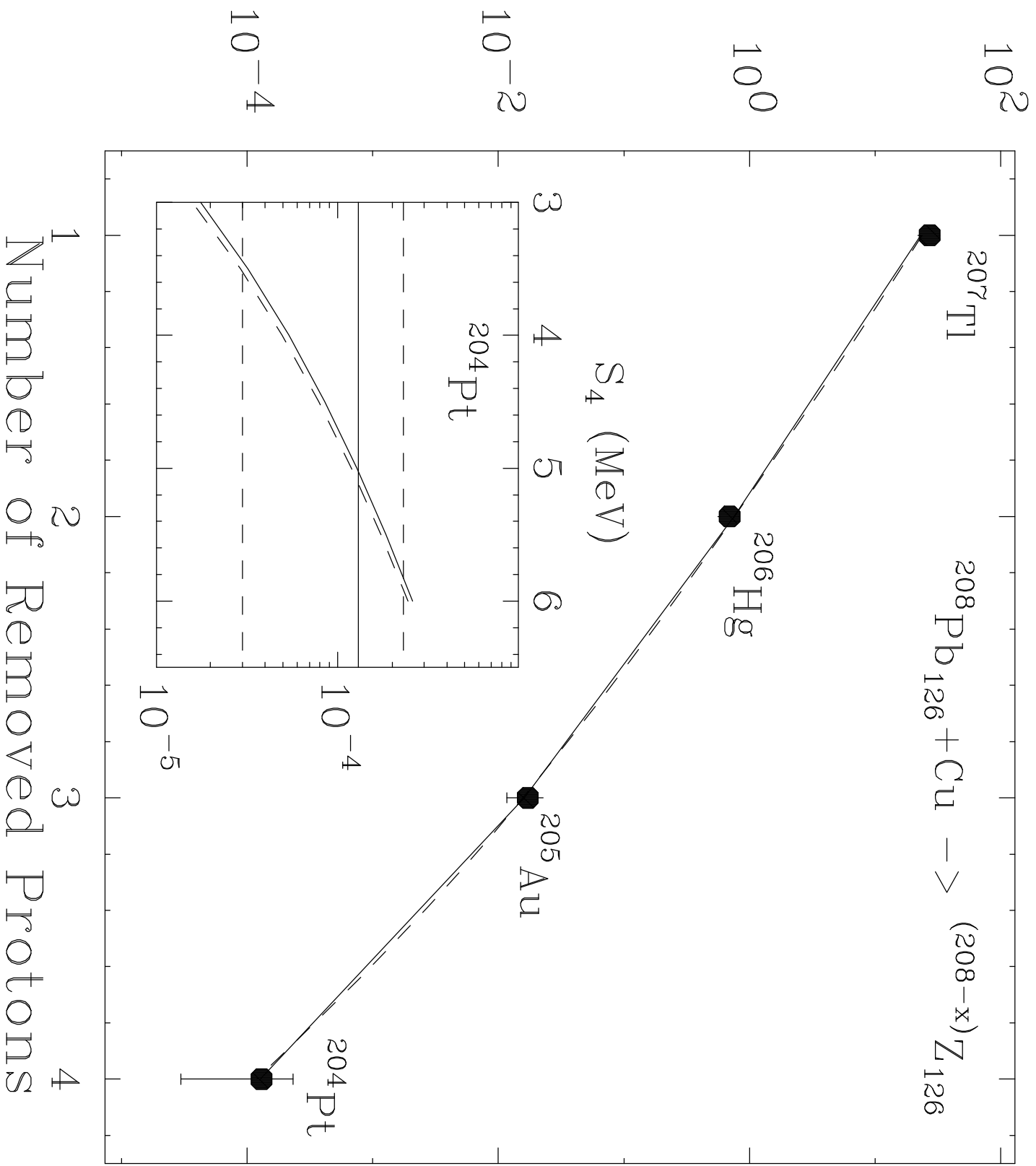
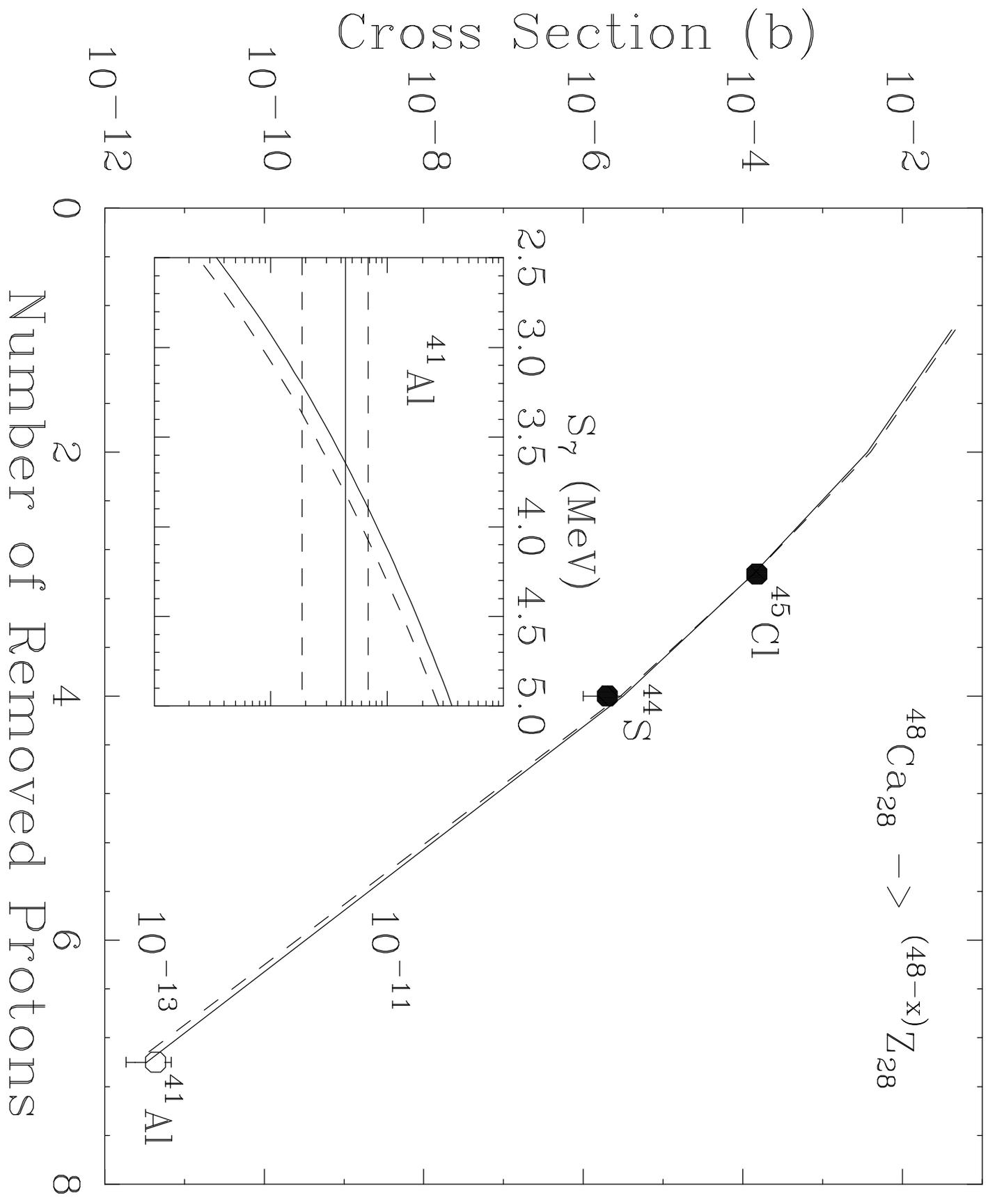
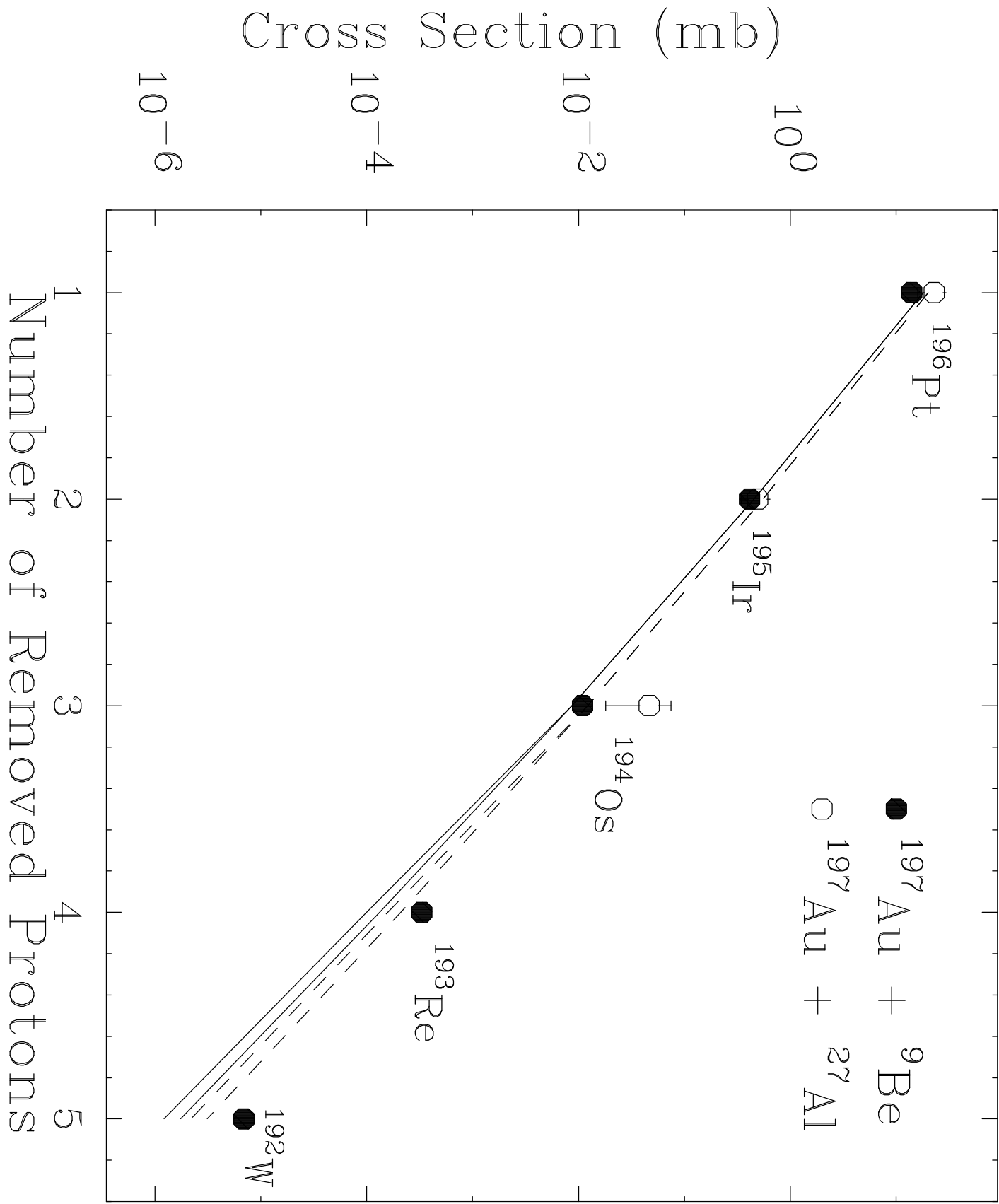
\end{document}